\documentclass[final,5p,times,twocolumn,sort&compress]{elsarticle}
\usepackage{color} 
\usepackage{amsmath}
\usepackage{graphicx}
\graphicspath{ {graphics/} }
\newcommand{\eq}[1]{Eq.~(\ref{#1})}
 \begin{document}
\begin{frontmatter}
  
  \title{Short range correlations and the isospin dependence of nuclear correlation functions}
  
  \author[mit]{R.~Cruz-Torres}
  \author[mit]{A.~Schmidt}
  \author[uwa]{G.~A.~Miller\corref{cor}}
  \ead{miller@uw.edu}
  \author[odu]{L.~B.~Weinstein}
  \author[huj]{N.~Barnea}
  \author[huj]{R.~Weiss}
  \author[tau]{E.~Piasetzky}
  \author[mit]{O.~Hen}
  
  \address[mit]{Massachusetts Institute of Technology, Laboratory for Nuclear Science, Cambridge, MA 02139, USA}
  \address[odu]{Old Dominion University, Department of Physics, Norfolk, VA 23529, USA}
  \address[tau]{Tel Aviv University, School of Physics and Astronomy, Tel Aviv 69978, Israel}
  \address[huj]{Hebrew University, Racah Institute of Physics, Jerusalem 91904, Israel}
  \address[uwa]{University of Washington, Department of Physics, Seattle, WA 98195, USA}
  
  \cortext[cor]{Corresponding author}
  
  \begin{abstract}
Pair densities and associated correlation functions provide a critical tool for introducing many-body correlations into 
a wide-range of effective theories. {\it Ab initio} calculations show that two-nucleon 
pair-densities exhibit strong spin and isospin dependence. However, such calculations 
are not available for all nuclei of current interest. We therefore provide a simple model, 
which involves combining  the short and long separation distance behavior using a single 
blending function,  to accurately describe the two-nucleon correlations inherent in existing 
{\it ab initio} calculations. We show that the salient features of the correlation function arise from the features of the two-body short-range nuclear interaction,
and that  the suppression of the $pp$ and $nn$ pair-densities caused by the Pauli principle is important.  Our procedure for obtaining pair-density functions and correlation functions can be applied to heavy nuclei which lack {\it ab initio} calculations.
  \end{abstract}
 
  \begin{keyword}
    correlation function, contact formalism, short range correlations
  \end{keyword}
  
\end{frontmatter}

\section{Introduction}

Correlation functions are a valuable tool for describing interacting many-body systems, 
providing a means of encapsulating complex many-body dynamics. In the absence of correlations,
a many-body probability density, such as that from a many-body quantum mechanical wave-function,
can be written as an anti-symmetrized product of single-particle probability densities. The correlation function
describes important deviations from this picture.  Our aim here is to explain the basic physics inputs that determine 
the nuclear pair-density functions and the correlation functions derived from them. This is done by blending the 
short-distance behavior, as determined by the contact formalism~\cite{Tan:2008,Weiss:2016obx, Weiss:2015mba}, with the 
known long distance behavior.  The input needed to use the contact formalism is  accessible from experimental data, as shown in 
Ref.~\cite{Weiss:2016obx}. 

Correlation functions are widely used in nuclear physics. For recent reviews see  Refs.~\cite{Hen:2016kwk,Atti:2015eda}. The nucleus
is a strongly-interacting, quantum mechanical, many-body system with high density and a complicated
interaction between constituent nucleons.  There is no fundamental central potential, so correlations must exist.
An early paper that modeled nuclear  correlation functions~\cite{Miller:1975hu} was used in a wide 
variety calculations (see the early review \cite{Haxton:1985am}) involving the strong and weak interactions, demonstrating the impact of correlation
functions on the field.  More recent examples in which correlation functions are crucial ingredients include: calculations of neutrinoless double beta 
decay~\cite{Kortelainen:2007rn,Kortelainen:2007rh,Kortelainen:2007mn,Menendez:2008jp,Simkovic:2009pp,Engel:2009gb},
nuclear transparency in quasielastic scattering~\cite{Mardor:1992sb,Frankel:1992er,Kundu:1999gx,Kundu:1998ti,Lee:1992rd,PhysRevC.77.034602},
shadowing in deep inelastic scattering~\cite{PhysRevC.52.1604}, and parity violation in 
nuclei~\cite{Adelberger:1985ik, Towner:1987zz}.  

Despite the wide use of correlation functions, their spin and isospin dependence has received less attention. 
  The nucleon-nucleon interaction is both spin and isospin
dependent, and these dependencies become very important at short-range, leading to phenomena such as
the strong preference for proton-neutron short-range correlated pairs~\cite{Tang:2002ww,
PhysRevLett.97.162504, PhysRevLett.99.072501, Subedi1476, PhysRevLett.113.022501, Hen614,Sargsian:2012sm}. 

\begin{figure}[htpb]
\includegraphics[width=\columnwidth]{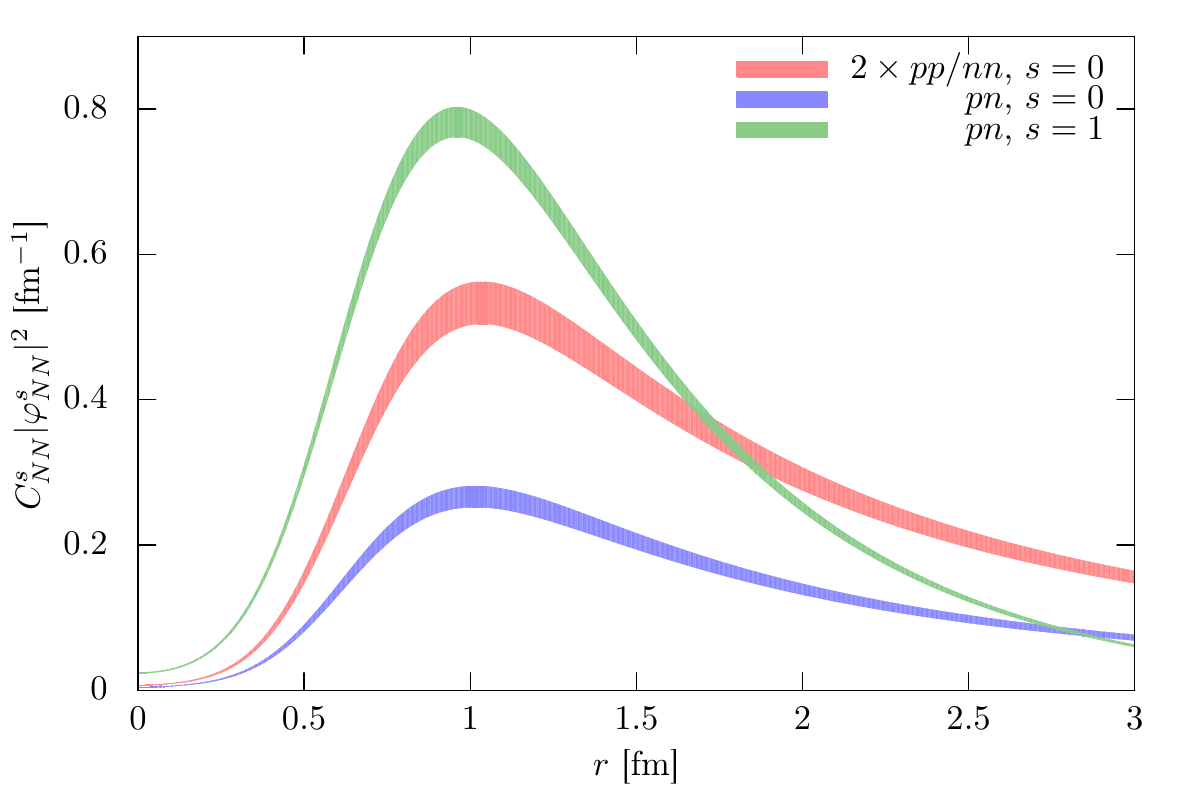}
\caption{\label{fig:contacts} In the two-body density from contact formalism~\cite{Weiss:2016obx, Weiss:2015mba},
the $np$ two-body density is dominated by spin-1 pairs. $^{40}$Ca, shown here, illustrates this universal behavior. 
For $r\le 0.9$ fm, these results reproduce those of Cluster Variational Monte Carlo (CVMC)~\cite{Pieper:1992gr}  calculations.
The $pp/nn$ spin-0 density is enhanced by a factor of 2 to provide some separation from $np$ spin-0.}
\end{figure}

The calculations in this paper use the formalism of nuclear contacts~\cite{Weiss:2016obx, Weiss:2015mba} 
to determine the spin and isospin decomposition of the two-body density that determines the correlation function. This formalism is 
based on the separation of scales inherent in the long- and short-range structure of nuclei~\cite{Weiss:2016obx, Weiss:2015mba}.
At short distances, the aggregate effect of long-range interactions can be encapsulated into coefficients, 
called ``contacts,'' which are nucleus-specific, while the underlying short-range behavior is a universal 
property of the two-body nuclear interaction. In the contact formalism, the two-body density, $\rho_{NN,s}(r)$, 
defining the probability for finding a nucleon-nucleon pair with separation distance $r$, can be modeled at
short distance ($r\lesssim 1$~fm) by:
\begin{equation}
\rho^\text{contact}_{NN,s}(r) = C_{A} ^{NN,s}\times |\varphi_{NN,s}(r)|^2\label{contact}
\end{equation}
for nucleus, $A$, where $C_{A}$ is the contact coefficent, ${NN}$ stands for proton-proton ($pp$), 
proton-neutron ($pn$), or neutron-neutron ($nn$) pairs and the index $s$ denotes the  spin $0,1$ 
of the two-nucleon systems. The wave functions $\varphi_{NN,s}(r)$ are zero-energy (S- or S-D wave) solutions to the
Schr\"{o}dinger equation with a modern nucleon-nucleon potential, e.g., AV18~\cite{Wiringa:1994wb}. 
Equation \ref{contact} assumes angle averaging, and the zero-energy nature restricts the number of
contacts. The key assumption in this formalism is that these functions, $\varphi_{NN,s}(r)$ can be used for
all nuclei. Contact coefficients can be determined for the different possible spin and isospin configurations 
of a nucleon-nucleon pair from experiment or from fitting {\it ab initio} calculations. Previous 
studies~\cite{Weiss:2016obx}, show that the $NN$ state with deuteron quantum numbers is dominant: 
the peak value of the product $C_{A}^{np,s=1}|\varphi_{np,s=1}(r)|^2$ is four times larger
than for any other combination. This dominance is caused by the tensor 
force~\cite{Alvioli:2007zz,Schiavilla:2006xx,Sargsian:2005ru} As an example, the decomposition of the 
two-body density from contact formalism for $^{40}$Ca is shown in Fig.~\ref{fig:contacts}.

\section{Describing the pair (two-body) density }

The two-body pair density distribution $\rho_{NN,s}(\vec{r})$, is defined as the
probability density for finding a nucleon-nucleon pair separated by $\vec{r}$, with relative spin $s$,
normalized so that its integral is the number of possible $NN,s$ pairs. The two-body density is expressed  as a matrix element of the 
nuclear wave function $|\psi\rangle$ by 
\begin{equation}
\rho_{NN,s}(\vec{r}) \equiv \sum_{\substack{i,j \in NN \\i<j}} \langle \psi | \delta(\vec{r} - \vec{r}_{ij})P_s| \psi \rangle, \label{2d}
\end{equation}
where  $\vec{r}_{ij}$ is the separation
between nucleons $i$ and $j$ and $P_s$ is a projection operator onto  the spin $s$ of the  nucleon pair.


Our aim here is to  provide a  
simple understanding of the underlying mechanisms that produce the isospin dependence and other features. 
We will compare our results for  $\rho_{NN}(r)$
to {\it ab initio} calculations performed using Cluster Variational Monte Carlo (CVMC)~\cite{Pieper:1992gr} 
of $^{16}$O and $^{40}$Ca, the two heaviest nuclei studied so far using CVMC~\cite{Lonardoni:2017egu}. 
Several other calculations that include the necessary spin and isospin dependence in computing densities are those of
 Refs.~\cite{Alvioli:2005cz,Alvioli:2007vv,Alvioli:2009ev,Feldmeier:2011qy,Pieper:1992gr,Alvioli:2007zz}.  A nice {\it ab initio} 
treatment of light nuclei has recently appeared~\cite{Pastore:2017ofx}. See also Ref.~\cite{Benhar:2014cka}, which 
is based on nuclear matter calculations.

To achieve the desired understanding we design a model in which the two-body density is formed from a 
combination of the correlated density coming from nuclear contact formalism (Fig.~\ref{fig:contacts}), 
which accounts for the behavior for $r\le 0.9$ fm and a longer-ranged term, $\rho_{NN}^\text{(0)}(r),$ 
for which correlations are expected to be unimportant. We define this term as  
$\rho_{NN}^\text{(0)}(r)$,  given by 
\begin{equation}
\rho_{NN}^\text{(0)}(\vec{r}) \equiv S_{NN} \int \text{d}^3\vec{R}\rho_N(\vec{R}+\vec{r}/2) \rho_{N}(\vec{R} - \vec{r}/2),
\label{eq:rho2_uncorrelated}
\end{equation}
where $\rho_N$ is the one-body density, normalized to proton or neutron number,  $\vec{R}$ represents 
the center-of-mass position of a nucleon-nucleon pair, and $S_{NN}$ represents a symmetry factor, which 
equals $1$ for $pn$ pairs, equals $Z(Z-1)/2Z^2$ for $pp$ pairs---since there are only $Z(Z-1)/2$ unique 
$pp$ pairs in a nucleus---and equals $N(N-1)/2N^2$ for $nn$ pairs. 

Then the full two-body density combines the short and long distance behavior, with the relative weighting
determined by a blending function, $g_{NN}(r)$, and constant, $\kappa$, such that
\begin{equation}
\rho_{NN}(r) = g_{NN}(r) \rho_{NN}^\text{contact}(r) + \kappa (1 - g_{NN}(r)) \rho_{NN}^\text{(0)}(r),
\label{eq:g_func}
\end{equation}
We can understand how the correlated and uncorrelated densities contribute to produce the specific
behavior of the correlation function seen through CVMC by assessing the quality of this model and 
by determining the blending function.

In order to parameterize $g_{NN}(r)$, we consider the short- and long-range constraints.
At short-distance, where $\rho_{NN}^\text{contact}(r)$ is an accurate description of the two-body 
density~\cite{Weiss:2016obx}, $g_{NN}(r)$ equals 1. For large distances, $\rho_{NN}$ must 
approach $\rho_{NN}^\text{uncorr.}$. Since $\rho_{NN}^\text{contact}$ falls off approximately as $1/r^2$ 
for $r>2$~fm, $g_{NN}$ must approach $(\kappa - 1 )/\kappa$ in the long-range limit, in order that the 
pair density approach $\rho_{NN}^\text{(0)}$. We propose the following model which meets these requirements:
\begin{equation}
g_{NN}(r) = \begin{cases}
1 & r \leq 0.9\text{~fm} \\
\frac{1}{\kappa} \left( \kappa - 1 + e^{(0.9\text{~fm} - r)/a}\right) & r > 0.9\text{~fm}
\end{cases},
\label{eq:g_def}
\end{equation}
For  $r<0.9$~fm, $\rho_{NN}(r)$ is modeled well by the contact expression~\eq{contact}
(see~\cite{Weiss:2016obx}). For $r>0.9$~fm, the contact 
density and the uncorrelated densities are blended, with a characteristic length-scale, $a$. 
In principle, $a$ would depend on the isospin of the pairs and on the specific nucleus being studied.
\begin{figure}[htpb]
\includegraphics[width=\columnwidth]{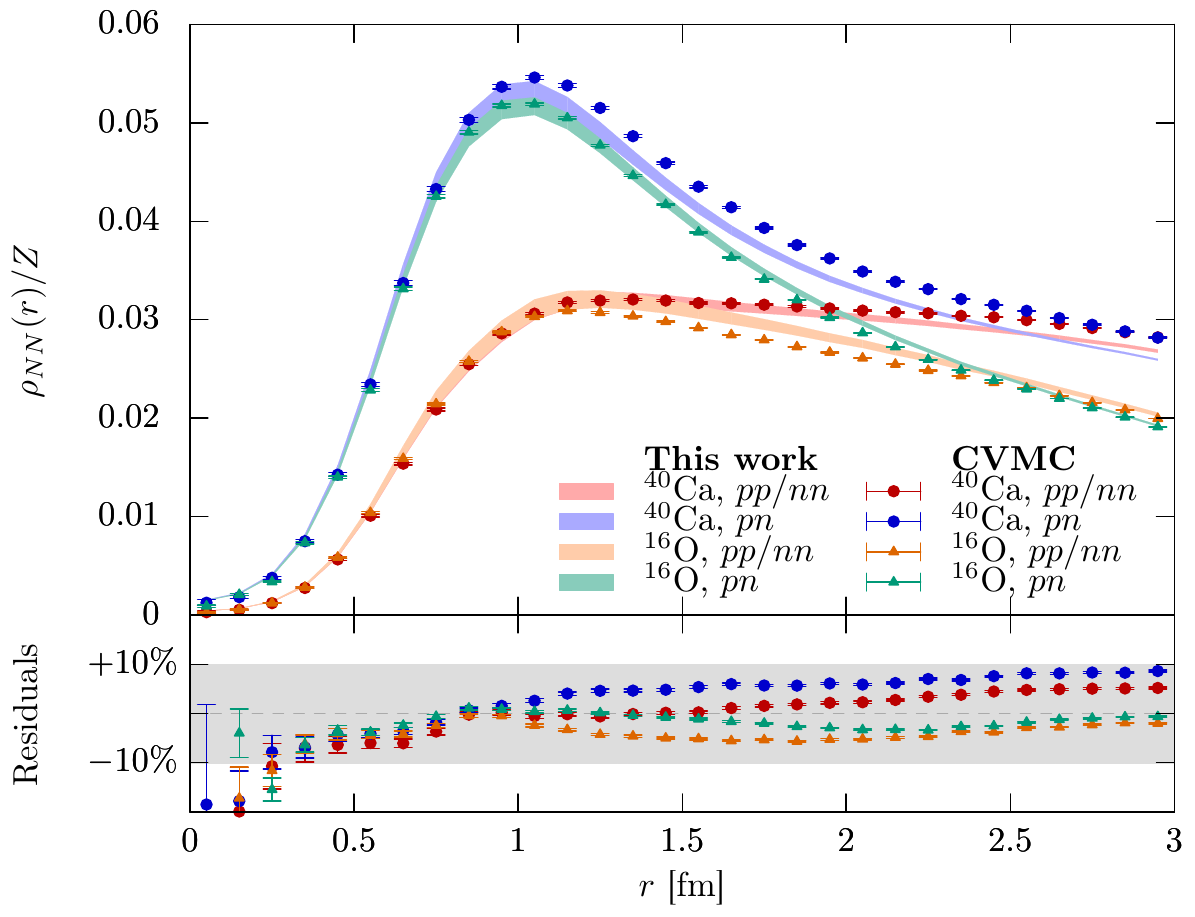}
\caption{\label{fig:rho2b} The model of  equation (\ref{eq:g_func}), with only a single fitted parameter,
  can reproduce the two-body densities  for both $pp$ and $pn$ pairs, and for both $^{16}$O and $^{40}$Ca,
  to within $\pm 10\%$. The results here are shown for $\kappa=2$. }
\end{figure} 

Varying the parameters of \eq{eq:g_def} to describe $pp$, $nn$ and $pn$ pairs in $^{16}$O and $^{40}$Ca
shows that the same blending function $g(r)$ can be used to describe all the two-body densities
calculated using CVMC, shown in Fig.~\ref{fig:rho2b}. CVMC correlation functions are shown
as points, while our model, described in equation \ref{eq:g_func}, is shown with bands, for which the
dominant contribution to the uncertainty comes from the contact coefficients, $C_{NN}$. The uncorrelated
density, $\rho_{NN}^\text{(0)}$, used by our model is supplied by CVMC calculations of the one-body
density $\rho_N$. The residuals show the difference between the CVMC density and those of
the model, divided by the model, with the error bars showing the uncertainties in the CVMC densities.
Our model is able to reproduce the correlation functions for both $pp$ and $pn$ pairs in
two different nuclei (As these CVMC calculations treat $p$ and $n$ symmetrically,
and since $^{16}$O and $^{40}$Ca are both symmetric nuclei, the results for $pp$ and $nn$ pairs are the same).
In achieving this description we find that the parameter $a$ depends smoothly on
$\kappa$. With $\kappa=2$, $a=1.518 \pm0.001$~fm. Fig.~\ref{fig:rho2b} shows that the simple  model qualitatively reproduces
CVMC calculations.

Fig.~\ref{fig:rho2b} demonstrates that the spin-isospin dependence of the two-body density
function   occur at short distances, and therefore originate 
  from the contact densities of
\eq{contact}, while the long range behavior is universal between different kinds of pairs 
and in different nuclei.

\section{Correlation Function}

The standard procedure for defining a  
 correlation function, $F_{NN,s}(r)$, a function of the separation distance between nucleons 
$r \equiv |\vec{r}|$, is to take the ratio of the fully correlated to the   two-body densities computed in the absence of dynamical correlations, i.e.,
\begin{equation}
F_{NN,s}(r) \equiv \frac{\rho_{NN,s}(r)}{\rho_{NN}^\text{uncorr.}(r)}.
\label{eq:F}
\end{equation}
The notation, $F_{NN,s}(r)$, is meant to convey that there can be differences in correlations between different
spin and isospin configurations. In cases where we refer to a generic correlation function, we will suppress the
indices and use $F(r)$.  The denominator must be treated with more sophistication than the function $\rho_{NN}^{(0)}$ 
used in the phenomenological fit presented above. The correlative effects of the Pauli principle must be included.

Typical applications of correlation functions in nuclear physics begin with anti-symmetrized wave functions, in the 
form of a Slater determinant. Using a Slater determinant, one can compute the uncorrelated two-body density as the 
matrix element of the two-body density operator. The result is 
\begin{multline}
    \rho_{NN}^\text{uncorr.}(\vec{r}) =
    \frac{1}{2}\sum_{\alpha,\beta,\in {\rm occ}} \int \text{d}^3r_1\text{d}^3r_2\delta(\vec{r}-(\vec{r}_1-\vec{r}_2)) 
    \phi_\alpha^\dagger(x_1)\phi_\beta^\dagger(x_2) \\
    \times [\phi_\alpha(x_1)\phi_\beta(x_2)-\phi_\beta(x_1)\phi_\alpha(x_2)],
\end{multline}
where $x_i$ represents several quantum numbers: $x\equiv(\vec{r},m_s=\pm1/2,m_t=\pm1/2)$.
{For the case of proton-neutron pairs, this reduces to the expression of Eq.~(\ref{eq:rho2_uncorrelated}).
However, for the case of two protons, one finds:}
\begin{align}
  \begin{split}  
    \rho_{pp}^\text{uncorr.}(\vec{r}) =& 
        {1\over 2}\int \text{d}^3r_1\text{d}^3r_2\delta(\vec{r}-(\vec{r}_1-\vec{r}_2))\\ 
        &\times \left[ \rho(\vec{r}_1)\rho(\vec{r}_2) -{1\over 2} \rho(\vec{r}_1,\vec{r}_2) \rho(\vec{r}_2,\vec{r}_1)\right],
  \end{split}\\
  \equiv & {Z\over Z-1}\rho_{pp}^\text{(0)}(\vec{r})-\rho_{pp}^\text{exch.}({\vec{r}}),
  \label{eq:den}
\end{align}
where $\rho(\vec{r})$ is the proton one-body density, normalized to $Z$. The expression is the same for
neutron-neutron pairs, substituting $N$ for $Z$ and the neutron one-body density for the proton one-body density. 
The quantity $\rho(\vec{r}_1,\vec{r}_2)$ is the density-matrix defined such that its diagonal elements yield the proton or 
neutron one-body density. The second term of \eq{eq:den} represents the influence of the Pauli exclusion principle: 
two spin-up protons cannot occupy the same orbital. This term is absent for the neutron-proton two-body density.
 
It is useful to avoid using a specific Slater determinant, which would depend on the nucleus. 
Instead, we  apply a result based on  nuclear matter (but using a local-density approximation) expressed as
\begin{equation}
\rho_{NN}^\text{exch.}(\vec{r}) = \frac{Z}{2(Z-1)}\rho_{pp}^\text{(0)}(r) \times \left( \frac{3 j_1(\bar{k}_F r)}{\bar{k}_F r} \right)^2,
\label{dme}
\end{equation}
where $\bar{k}_F$ is a Fermi momentum (averaged over the nuclear volume) and $j_1$ is a spherical Bessel function. We use this approximation throughout, with
$\bar{k}_F$ assumed to be 200~MeV$/c$.This approximation amounts to using the local-density approximation to the first term of the density-matrix expansion
of Ref.~\cite{Negele:1972zp}.  We verify the  accuracy of \eq{dme} numerically, by comparing  with  the Slater determinant provided by the single-particle wave functions of  
 Ref.~\cite{Negele:1970jv}.
 
\begin{figure}[htpb]
\includegraphics[width=\columnwidth]{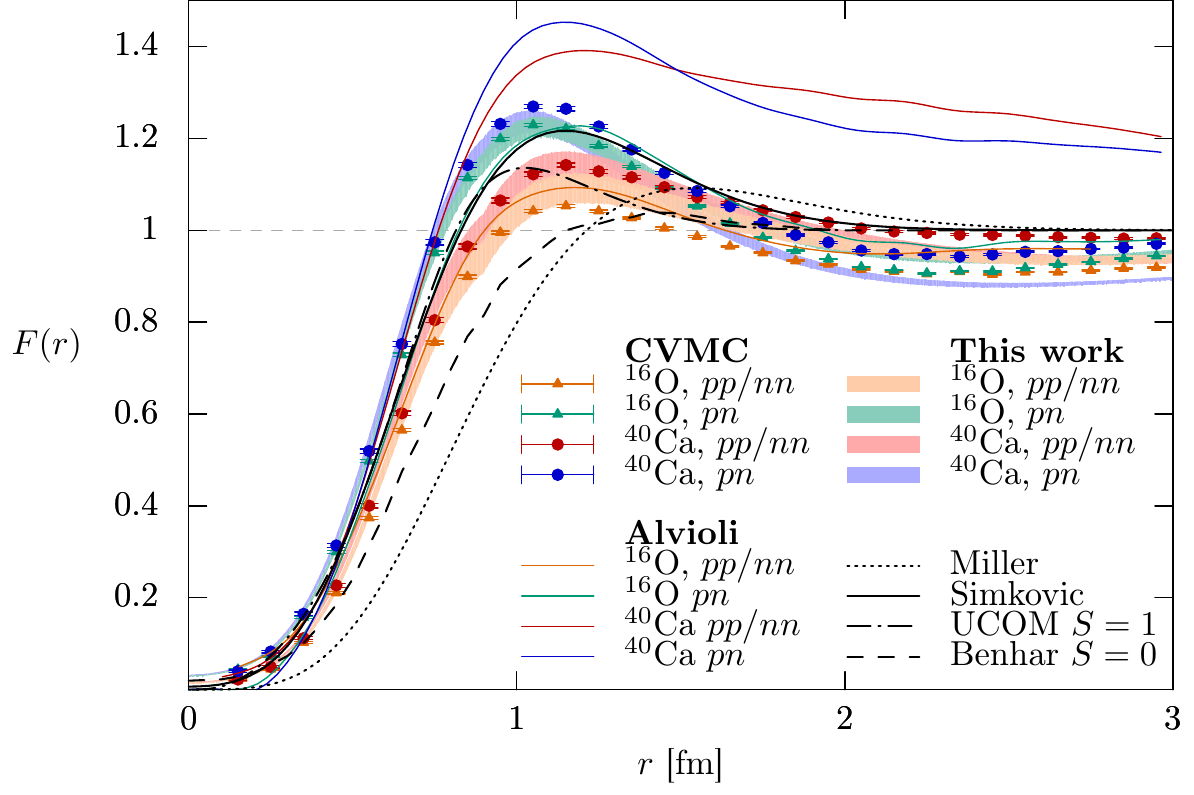}
\caption{\label{fig:money_plot} The model of  equation (\ref{eq:g_func}), with only a single fitted parameter,
  can reproduce the correlation functions for both $pp$ and $pn$ pairs, and for both $^{16}$O and $^{40}$Ca,
  to within $\pm 10\%$. The results here are shown for $\kappa=2$. The predictions of Miller
  and Spencer~\cite{Miller:1975hu}, Simkovic et al.~\cite{Simkovic:2009pp}, Alvioli et al.~\cite{Alvioli:2005cz},
  Benhar et al.~\cite{Benhar:2014cka}, and the UCOM calculation by Roth et al.~\cite{Roth:2005pd} 
  are shown for comparison.}
\end{figure}

The points in Fig.~\ref{fig:money_plot} show correlation functions calculated using equations~\ref{eq:rho2_uncorrelated},
\ref{dme}, and~\ref{eq:F}, with CVMC providing the one- and two-body densities. As can be seen, the correlation
functions are similar for $^{16}$O and $^{40}$C. But there is some isospin dependence, as displayed by 
the differences at $r<1.5$~fm between $pp$- and $pn$-pairs (dominated by $s=1$).  Note also that  because  these CVMC 
calculations treat $p$ and $n$ symmetrically, and since $N=Z$ for  both $^{16}$O and $^{40}$Ca, 
the results for $pp$ and $nn$ pairs are the same.

{

Fig.~\ref{fig:money_plot} also shows, for comparison, several other calculations of nuclear correlation
functions, including the original model suggested by Miller and Spencer \cite{Miller:1975hu} as well as 
more recent work. The correlation functions from {\it ab initio} and from our model are close to that of 
Simkovic et al.~\cite{Simkovic:2009pp} and to the $^{16}$O calculations of Alvioli et al.~\cite{Alvioli:2005cz},
but are higher than the correlation functions predicted by Benhar et al.~\cite{Benhar:2014cka} and by 
Miller and Spencer. The calculations by Alvioli et al.~for $^{40}$Ca predict a significantly higher correlation
function for both $pp/nn$ and $pn$. A calculation using the Unitary Correlation Operator Method 
(UCOM)~\cite{Roth:2005pd} in the $T=0$, $S=1$ channel is slightly lower than our predictions for $pn$ pairs. 
A study of Jastrow correlation functions ~\cite{Engel:2011ss} is relevant in the present context.
These comparisons show that at the level of two-body cluster truncation, isospin symmetry is broken, 
and that the Miller-Spencer parameterization suffers from this problem. The Simkovic et al.\ model avoids
this problem because of the bump (at about $r=$ 1 fm) in their correlation function. Our present reproduction of the
correlation function of~\cite{Simkovic:2009pp} shows that our work also avoids this problem. It is necessary 
to keep the effects of the Pauli principle in mind when making comparisons between correlation functions 
produced by different authors.

}

For convenience, we provide the following parameterization for the $pp/nn$ and $pn$ correlation functions
determined from CVMC:
\begin{equation}
F(r)= 1 - e^{-\alpha r^2} \times \left(\gamma + r \sum\limits_{i=1}^3 \beta_i r^i \right)
\label{eq:Fform}
\end{equation}
with parameter values given in table \ref{table:Fparams}.  This function reproduces the correlation functions
of both $^{16}$O and $^{40}$Ca. 

\begin{table}
\caption{\label{table:Fparams} Parameters describing $F(r)$, using the functional form of equation \ref{eq:Fform}}
\centering
\begin{tabular}{ c c c c }
\hline \hline
Parameter & Units & Value ($pp/nn$) & Value ($pn$)\\
\hline
$\alpha$ & fm$^{-2}$ & 3.17 & 1.08 \\
$\gamma$ & -- & 0.995 & 0.985 \\
$\beta_1$ & fm$^{-2}$ & 1.81 & -0.432 \\
$\beta_2$ & fm$^{-3}$ & 5.90 & -3.30 \\
$\beta_3$ & fm$^{-4}$ & -9.87 & 2.01 \\
\hline
\hline
\end{tabular}
\end{table}

\begin{figure}[htpb]
\includegraphics[width=\columnwidth]{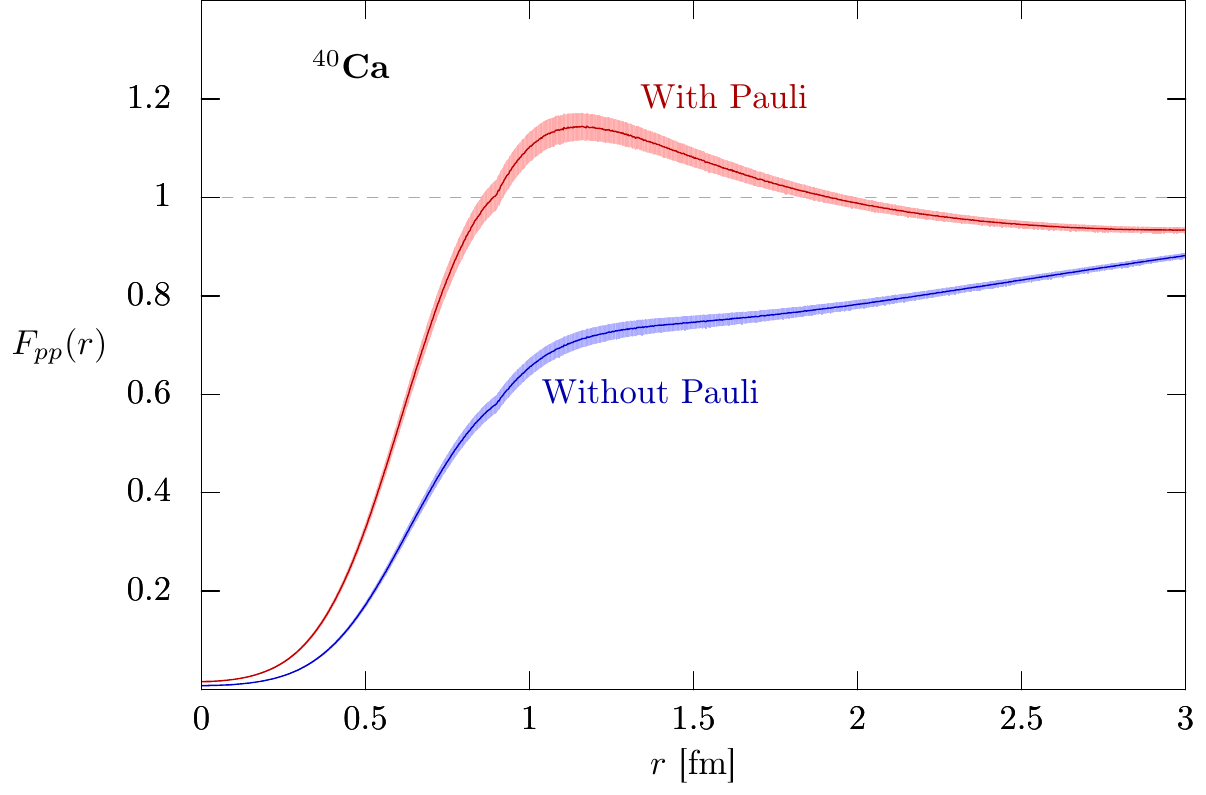}
\caption{\label{fig:pauli}
  Pauli exchange has a significant effect on $pp$ and $nn$ correlations. Correlations taken relative to
  a classical uncorrelated density (blue) (\eq{eq:rho2_uncorrelated}) appear significantly suppressed compared to correlations taken
  relative to an uncorrelated density that includes Pauli exchange (red). 
}
\end{figure}

Note that the Figs.~\ref{fig:contacts} and  \ref{fig:money_plot} display a striking contrast. The huge differences between like and un-like nucleon pairs seen in the former figure do not show up in the latter figure. This is because of the influence of the Pauli principle (as manifest in the second term of \eq{eq:den}, which strongly enhances the $pp$ and $nn$ correlation functions defined in \eq{eq:F}. This effect is displayed in Fig.~\ref{fig:pauli}.

\section{Discussion}

Our model,  defined in \eq{eq:g_func} and \eq{eq:g_def},  for the pair-density (\eq{2d}) requires the minimal  input of  the blending function, the nuclear contact coefficients and the universal functions  $|\varphi_{NN,s}(r)|^2$ of \eq{contact}. The   isospin-dependence  of the two-body pair density is produced by 
short-ranged interactions, driven by the two-nucleon tensor force.  While the nuclear contacts used in this work were determined
from CVMC calculations, they can also be determined from experimental data, as shown in 
Ref.~\cite{Weiss:2016obx}.   Thus one may obtain the pair-density for heavy nuclei for which  {\it ab initio} calculations do not exist.

The correlation function, \eq{eq:F}, can also be obtained. One needs  
  a one-body 
density function  to form $\rho_{NN}^\text{(0)}(r)$. The effects of the Pauli principle must be included as (for example)  in \eq{dme}.
One-body
densities have been well-measured experimentally, and simple parameterizations exist for many different
nuclei, e.g., Ref.~\cite{DeJager:1987qc}. 
 
In summary, we have provided a procedure  that  enables predictions of  two-body densities and correlation functions for nuclei that 
are too large for adequate {\it ab initio} calculations. 

\section{Acknowledgements}

We wish to thank Mark Strikman, Misak Sargsian, Claudio Ciofi degli Atti,  Max Alvioli, and Jan Ryckebusch for helpful
discussions. This work was supported by the U. S. Department of Energy Office of Science, Office of Nuclear Physics under 
Award Numbers DE-FG02-97ER-41014, DE-FG02-94ER40818 and DE-FG02-96ER-40960, the Pazy Foundation, and by the 
Israel Science Foundation (Israel) under Grants Nos. 136/12 and 1334/16. GAM thanks MIT LNS for and the Argonne National Laboratory for their hospitality during 
completion of this work.

\bibliographystyle{model1-num-names}

\bibliography{references.bib}

\end{document}